\documentclass[aps,twocolumn,floatfix,altaffilletter,superscriptaddress,preprintnumbers,
tightenlines,showpacs,showkeys,notitlepage,nofootinbib,
longbibliography]{revtex4-2}

\usepackage{amsmath,amssymb}
\usepackage{graphicx}
\usepackage{tabularx}
\usepackage{fancybox}
\usepackage[export]{adjustbox}
\usepackage{url}
\usepackage[dvipsnames]{xcolor}
\usepackage{siunitx}
\usepackage{orcidlink}
\usepackage{titlesec}
\usepackage{ulem}

\usepackage{microtype}
\usepackage{dcolumn}
\usepackage{bm,bbm}
\usepackage{multirow}
\usepackage{afterpage}

\titleformat*{\section}{\centering\bfseries\MakeUppercase}
\titlelabel{\thetitle\quad}
\titleformat*{\paragraph}{\bfseries}
\titlespacing*{\paragraph}{0pt}{3.25ex plus 1ex minus .2ex}{1em}

\usepackage{hyperref}
\usepackage[capitalize]{cleveref}
\newcommand\myshade{80}
\colorlet{mylinkcolor}{ForestGreen}
\colorlet{mycitecolor}{Red}
\colorlet{myurlcolor}{violet}
\hypersetup{
	linkcolor  = mylinkcolor!\myshade!black,
	citecolor  = mycitecolor!\myshade!black,
	urlcolor   = myurlcolor!\myshade!black,
	colorlinks = true
}

\newcommand{\dcp}{\delta_{\rm CP}}
\newcommand{\nova}{NO$\nu$A~}

\newcommand{\tx}{{\theta_{12}}}
\newcommand{\ty}{{\theta_{13}}}
\newcommand{\tz}{{\theta_{23}}}
\newcommand{\dl}{{\Delta_{31}}}
\newcommand{\ds}{{\Delta_{21}}}

\newcommand{\ahat}{\hat{A}}
\newcommand{\dhat}{\hat{\Delta}}

\newcommand{\datm}{\Delta m^2_{31}}
\newcommand{\dsun}{\Delta m^2_{21}}

\newcommand{\pmumu}{P(\nu_\mu \rightarrow \nu_\mu)}
\newcommand{\pme}{P_{\mu e}}

\newcommand{\pmebar}{P_{\bar{\mu} \bar{e}}}

\def\nue{{\nu_e}}
\def\anue{{\bar\nu_e}}
\def\numu{{\nu_{\mu}}}
\def\anumu{{\bar\nu_{\mu}}}

\newcommand{\cf}{{\it cf.}}

\newcommand{\be}{\begin{equation}}
	\newcommand{\ee}{\end{equation}}
\newcommand{\bea}{\begin{eqnarray}}
	\newcommand{\eea}{\end{eqnarray}}

\begin{document}
	
	
	\title{Diagnosing Unmodeled Neutrino Physics via DUNE and T2HK Complementarity}

	\newcommand{\Lee}{State Key Laboratory of Dark Matter Physics, Tsung-Dao Lee Institute \& School of Physics and Astronomy, Shanghai Jiao Tong University, Shanghai 200240, China}
	\newcommand{\Leeb }{Key Laboratory for Particle Astrophysics and Cosmology (MOE) \& Shanghai Key Laboratory for Particle Physics and Cosmology, Shanghai Jiao Tong University, Shanghai 200240, China}
	
	\newcommand{\tifr}{Tata Institute of Fundamental Research, Homi
		Bhabha Road, Colaba, Mumbai 400005, India}
	
	\author{Jo\~ao Paulo Pinheiro
		\orcidlink{0000-0002-8777-9338}\,}
	\email{joaopaulo.pinheiro@sjtu.edu.cn}
	\affiliation{\Lee}
	\affiliation{\Leeb}
	
	\author{Ushak Rahaman
		\orcidlink{0000-0003-0419-2970}\,}
	\email{ushak.rahaman@cern.ch}
	\affiliation{\tifr}
	
	\date{\today}
	
	\begin{abstract}
		\noindent
		Unmodeled beyond Standard Model (BSM) physics in neutrino 
		propagation can masquerade as parameter degeneracies in 
		future precision measurements. Because the upcoming DUNE 
		and T2HK experiments will operate at substantially 
		different baselines, interpreting their data under the 
		standard three-flavor framework provides a powerful 
		diagnostic tool: any propagation BSM effect can manifest as an artificial tension between their 
		extracted parameters. We demonstrate this principle using 
		the complex non-standard interactions (NSI) currently 
		favored to resolve the $\sim2\sigma$ tension between 
		NO$\nu$A and T2K. If these NSI solutions are realized, the 
		NSI-induced interference term $\propto\sin(\delta_{\rm CP}+\phi)$ 
		systematically distorts the DUNE appearance rates, leading 
		to a correlated misidentification of the atmospheric 
		mixing octant and the CP phase $\delta_{\rm CP}$. 
		Specifically, for $\varepsilon_{e\mu}$ 
		($\varepsilon_{e\tau}$) NSI, the DUNE fit shifts toward CP-
		conserving values (the opposite CP half-plane) along with 
		a preference for the wrong octant. In contrast, the 
		shorter-baseline T2HK experiment remains largely 
		insensitive to this effect. The resulting $\sim3\sigma$ 
		incompatibility between the DUNE and T2HK standard-fit 
		results (after 6 years of data collection 
		for each experiment)
		can provide a powerful experimental diagnostic for 
		propagation NSI, illustrating how baseline complementarity 
		can be a valuable tool to uncover new physics in the precision era.
	\end{abstract}
	
	\maketitle

	\textbf{Introduction:} 
	Neutrino oscillation physics has entered the precision era. Recent
	results from JUNO~\cite{JUNO:2025gmd} have determined the solar
	oscillation parameters with unprecedented accuracy and, when combined
	with global oscillation data, show a preference for the normal mass
	ordering (NO)\cite{Esteban:2026phq,Esteban:2024eli}. Cosmological measurements of baryon acoustic
	oscillations by CMB\cite{Planck:2018vyg} and DESI~\cite{DESI2024, DESI2024b} further disfavour the inverted
	ordering. With the mass ordering picture becoming clearer, two key
	unknowns remain: the octant of the atmospheric mixing angle
	$\theta_{23}$ and the value of the Dirac CP phase $\dcp$.
	
	The upcoming Deep Underground Neutrino Experiment 
	(DUNE)~\cite{DUNE:2018tke} ($L=1300$~km) and the Tokai-to-Hyper-Kamiokande 
	(T2HK) experiment~\cite{Abe:2018uyc} ($L=295$~km) are designed to determine 
	these remaining parameters with unprecedented precision through the 
	$\nu_\mu\to\nu_e$ appearance channel. However, as we enter this high-precision 
	regime, unmodeled beyond Standard Model (BSM) physics in neutrino propagation 
	poses a critical challenge: it can easily masquerade as standard parameter 
	degeneracies. Because DUNE and T2HK operate at drastically different baselines, 
	their corresponding matter effects differ significantly. Consequently, any unmodeled 
	BSM propagation effect will modify their appearance probabilities differently. 
	Fitting their combined future data under the strict assumption of standard 
	three-flavor oscillations can generate an artificial tension between 
	the parameters inferred by each experiment. This baseline complementarity, 
	therefore, transforms the combination of DUNE and T2HK into a powerful diagnostic tool, generalized 
	diagnostic tool for new physics.
	
	To concretely demonstrate this diagnostic capability, we can look to the 
	current generation of long-baseline experiments for a highly motivated BSM 
	test case. The T2K~\cite{T2K:2001wmr} ($L=295$ km) and 
	NO$\nu$A~\cite{NOvA:2007rmc} ($L=810$ km) experiments currently provide the 
	leading measurements of the appearance channel, but they exhibit a mild but 
	persistent inconsistency under the assumption of normal ordering. T2K observes a
	large neutrino--antineutrino appearance asymmetry consistent with
	near-maximal CP violation ($\dcp\simeq-\pi/2$), whereas NO$\nu$A
	measures a smaller asymmetry than expected for this value of $\dcp$.
	Because matter effects enhance the asymmetry at NO$\nu$A relative to
	T2K under NO, the two results cannot be simultaneously accommodated
	within the standard three-flavor framework, leading to a $\sim2\sigma$ tension~\cite{NOvA:2021nfi, Wolcott:2024,T2Kapp,T2Kdisapp,T2K:2025wet}.
	
	A compelling explanation for this discrepancy is provided by neutral-current 
	non-standard interactions (NSI) during propagation 
	\cite{Wolfenstein:1977ue,Mikheyev:1985zog, Grossman:1995wx, Proceedings:2019qno}. 
	Complex off-diagonal NSI introduce additional phases in the matter potential, 
	generating a new interference term in the oscillation probability. Because this
	contribution grows with baseline and neutrino energy, it affects
	NO$\nu$A more strongly than T2K and can reconcile the two current data sets
	while preserving normal ordering\cite{Miranda:2019ynh,Yu:2024nkc,
		Chatterjee:2020kkm,Rahaman:2022rfp,Chatterjee:2024kbn,
		Denton:2020uda,Cherchiglia:2023ojf, Alonso-Alvarez:2024wnh}. 
	
	In this work, we adopt these specific propagation NSI solutions as a realistic 
	proxy for generic unmodeled new physics. We show that if these solutions are 
	realized in nature, a standard analysis of DUNE measurements simultaneously 
	misidentifies the octant of $\theta_{23}$ and the value of $\dcp$. The two 
	biases are correlated, arising from the same NSI-induced CP-odd interference 
	term in the appearance probability. In contrast, the shorter-baseline T2HK
	experiment is largely insensitive to this propagation effect.
	Consequently, the standard-fit results of DUNE and T2HK become
	incompatible, yielding a $3.3$--$3.4\,\sigma$ tension 
	in a parameter-goodness-of-fit test, 
	after 6 years of data collection for each experiment.
	This discordance illustrates exactly 
	how baseline complementarity can act as an experimental diagnostic for new 
	physics in the precision era. This complimentarity was studied in the context of standard oscillation physics in ref.~\cite{Agarwalla:2024kti}.
	\\
	\textbf{NOvA--T2K tension and NSI as a resolution:} 
	As outlined above, the current discrepancy between NO$\nu$A 
	and T2K serves as an ideal test case for how differing matter 
	effects can expose unmodeled propagation physics. To understand 
	the mechanics of this tension, we note that the most
	important observable governing 
	both $\dcp$ and the mass ordering in long-baseline experiments is 
	the matter--antimatter appearance asymmetry
	\begin{equation}
		A_{CP}\equiv P(\nu_\mu\to\nu_e)-P(\bar\nu_\mu\to\bar\nu_e).
		\label{eq:ACP}
	\end{equation}
	
	T2K observes a large positive asymmetry consistent with near-maximal
	CP violation ($\dcp\simeq-\pi/2$). Under normal ordering (NO),
	matter effects enhance this asymmetry at the longer-baseline
	NO$\nu$A experiment. Consequently, the standard framework predicts a
	larger asymmetry in NO$\nu$A if the T2K result is interpreted as
	$\dcp\simeq-\pi/2$. Instead, NO$\nu$A measures a smaller asymmetry,
	leading to the aforementioned $\sim2\sigma$ tension between the two experiments under
	NO~\cite{NOvA:2021nfi,Wolcott:2024,T2Kapp,T2Kdisapp,T2K:2025wet}.
	
\begin{figure}[t]
		\centering
		\includegraphics[width=\columnwidth]{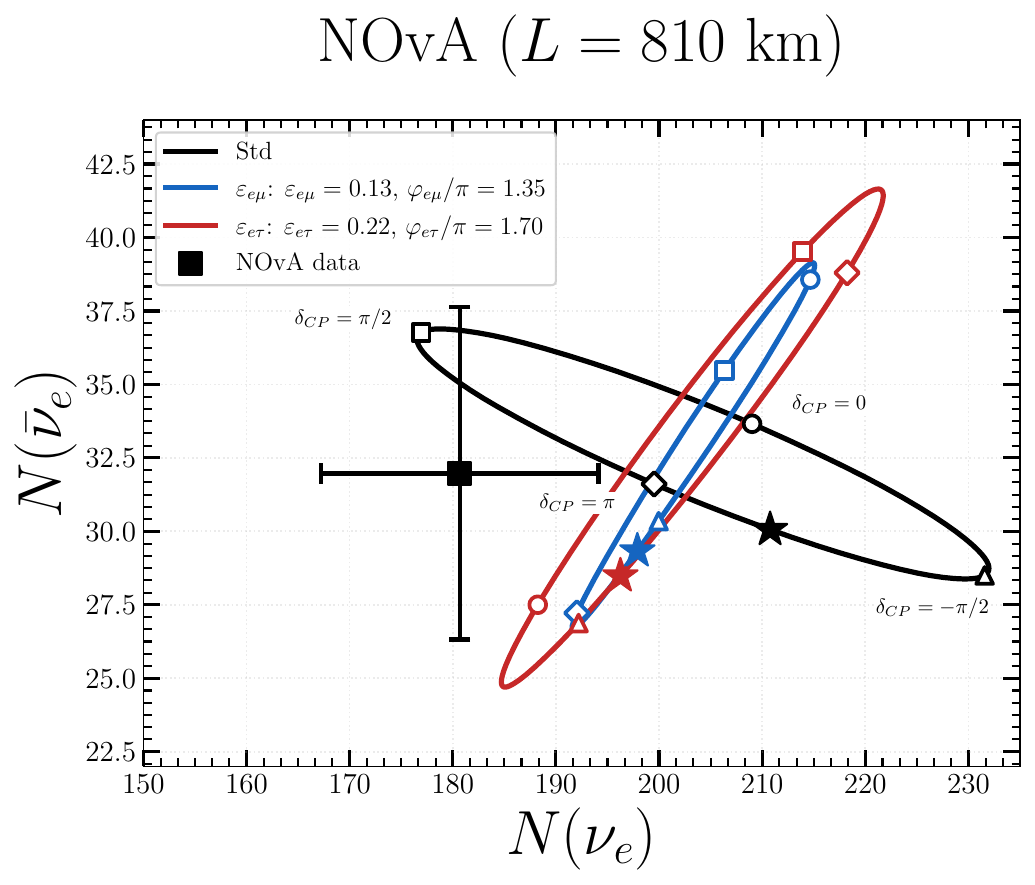}
		\includegraphics[width=\columnwidth]{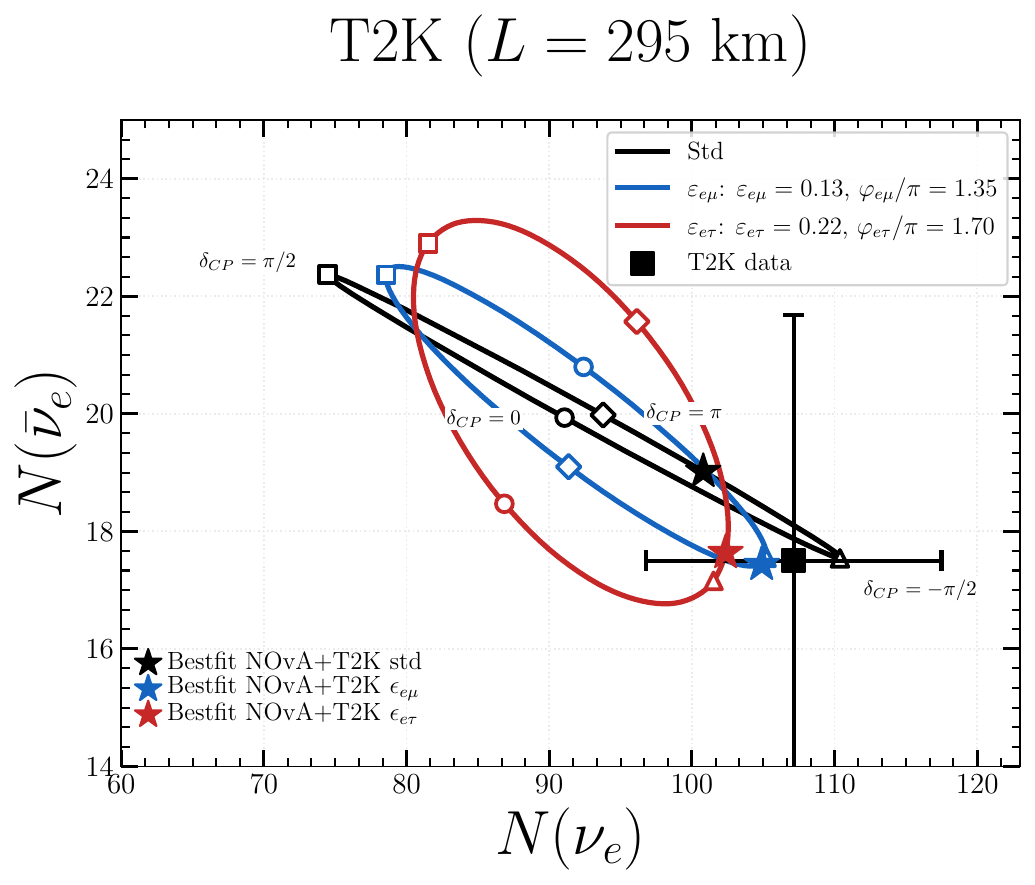}
		\caption{\footnotesize Bi-event plot for \nova (upper) and T2K
			(lower).  The ellipses are obtained by varying $\dcp$ over
			$[-\pi,\pi]$ with all other oscillation parameters fixed at the
			combined best-fit\cite{Esteban:2024eli}.  The data points are represented by 
			black squares and the predicted number of events are represented as
			stars. The circle, square, diamond and triangle shapes represents
			$\dcp=0,\pi/2,\pi(-\pi)$ and $-\pi/2$ respectively.}
		\label{bievents}
	\end{figure}

	This feature can be understood from the perturbative expansion of the
	$\nu_\mu\to\nu_e$ appearance probability in matter
	\cite{Cervera:2000kp,Freund:2001pn, Akhmedov:2004ny}. The standard probability can be
	written as $P_{\rm std}=P_0+P_1$, where the leading atmospheric term
	$P_0\simeq4s_{13}^2s_{23}^2f^2$ (with
	$f=\sin[(1-\ahat)\dhat]/(1-\ahat)$ a matter-modified oscillation
	function, the dimensionless term $\ahat=2EV_{\rm CC}/\dl$, where $V_{\rm CC}=\sqrt{2}G_F n_e$,
	with $G_F$ the Fermi constant and $n_e$ the electron density
	in Earth matter, and $\dhat=\dl L/(4E)$) carries no CP phase, while the
	interference term $P_1$ contains $\cos(\dhat+\dcp)$. Under
	$\nu\leftrightarrow\bar\nu$ both $\ahat$ and $\dcp$ change sign,
	yielding an asymmetry
	\begin{equation}
		A_{CP}^{\rm std}\approx
		16J_r\frac{\alpha}{\Delta_{31}}\dl\sin^2\!\dhat\,\sin\dcp
		+ A_{\rm mat}(\ahat,\dhat),
		\label{eq:ACP_SM}
	\end{equation}
	where $A_{\rm mat}$ denotes the purely matter-induced contribution,
	and $J_r=s_{12}c_{12}s_{13}c_{13}^2s_{23}c_{23}$ is the reduced Jarlskog
	factor (the full Jarlskog invariant being $J=J_r\sin\dcp$).
	For NO this term has the same sign as the CP contribution, so the
	predicted asymmetry at NO$\nu$A is generically larger than that at
	T2K. As illustrated in Fig.~\ref{bievents}, the standard bi-event
	ellipses therefore fail to simultaneously reproduce the observed
	event counts of the two experiments.
	A natural explanation of this discrepancy is provided by
	neutral-current non-standard interactions (NSI) in neutrino
	propagation
	\cite{Rahaman:2021leu,Miranda:2019ynh,Chatterjee:2020kkm,
		Rahaman:2022rfp,Chatterjee:2024kbn,Denton:2020uda,
		Cherchiglia:2023ojf}. Such effects arise from dimension-six operators
	\begin{equation}
		\mathcal{L}_{\rm NSI} =
		-2\sqrt{2}G_F\,\varepsilon_{\alpha\beta}^{fC}
		(\bar{\nu}_\alpha\gamma^\mu P_L\nu_\beta)
		(\bar f\gamma_\mu P_C f),
	\end{equation}
	which modify the matter potential in the flavor basis,
	\begin{equation}
		H = H_{\rm vac} + V_{\rm CC}
		\begin{pmatrix}
			1+\varepsilon_{ee} & \varepsilon_{e\mu} & \varepsilon_{e\tau}\\
			\varepsilon_{e\mu}^* & \varepsilon_{\mu\mu} & \varepsilon_{\mu\tau}\\
			\varepsilon_{e\tau}^* & \varepsilon_{\mu\tau}^* & \varepsilon_{\tau\tau}
		\end{pmatrix},
		\label{eq:Ham}
	\end{equation}
	with effective parameters
	\begin{equation}
		\varepsilon_{\alpha\beta}=
		\sum_{f\in\{e,u,d\}}
		\frac{n_f}{n_e}\varepsilon_{\alpha\beta}^f .
		\label{eq:eff_nsi}
	\end{equation}
	For the Earth's crust \cite{PREM},
	\begin{equation}
		\frac{n_u}{n_e}=2+Y_n\simeq3.05,
		\qquad
		\frac{n_d}{n_e}=1+2Y_n\simeq3.10.
		\label{eq:Yn}
	\end{equation}
	The off-diagonal couplings are generally complex,
	$\varepsilon_{\alpha\beta}=|\varepsilon_{\alpha\beta}|
	e^{i\phi_{\alpha\beta}}$.
	
	In the presence of $\varepsilon_{e\mu}$ or $\varepsilon_{e\tau}$,
	the appearance probability can be written as
	\begin{equation}
		\pme^{\rm std+NSI}=P_0+P_1+P_2 ,
		\label{pme-nsi}
	\end{equation}
	where the standard terms are
	\begin{eqnarray}
		P_0&\simeq&4s_{13}^2s_{23}^2f^2,
		\label{P0}\\
		P_1&\simeq&8s_{13}s_{12}c_{12}s_{23}c_{23}\alpha
		fg\cos(\dhat+\dcp),
		\label{P1}
	\end{eqnarray}
	and the NSI contribution is
	\begin{eqnarray}
		P_2&\simeq&8s_{13}s_{23}\ahat|\varepsilon|
		\left[a f^2\cos(\dcp+\phi) \right. \nonumber\\
		&&\left. +\, bfg\cos(\dhat+\dcp+\phi)\right],
		\label{P2}
	\end{eqnarray}
	where $\alpha=\frac{\ds}{\dl}$, and $g=\frac{\sin(\ahat\dhat)}{\ahat}$. The NSI coupling is parametrized as
	$\varepsilon_{e\mu}=|\varepsilon_{e\mu}|e^{i\phi_{e\mu}}$, with $a=\sin^2\theta_{23}$ and
	$b=\cos^2\theta_{23}$.
	For $\varepsilon_{e\tau}=|\varepsilon_{e\tau}|e^{i\phi_{e\tau}}$,
	the structure of Eq.~\ref{P2} remains the same, but the
	coefficients become $a=s_{23}c_{23}$ and $b=-s_{23}c_{23}$,
	while all other terms remain unchanged. The antineutrino
	probability $\pmebar$ is obtained by reversing the signs of
	$\ahat$, $\dcp$, and $\phi$. 
	The new term $P_2$ modifies both the total appearance probability and
	the neutrino–antineutrino asymmetry. In particular, the asymmetry
	receives an additional contribution
	\begin{equation}
		\Delta A_{CP}^{\rm NSI}
		\propto \ahat\,a\,|\varepsilon|\,\sin(\dcp+\phi),
		\label{eq:ACP_NSI}
	\end{equation}
	introducing an effective CP phase $(\dcp+\phi)$.
	Because $P_2\propto\ahat|\varepsilon|$, the NSI effect grows with
	baseline and energy and therefore impacts NO$\nu$A more strongly than
	T2K. As illustrated in Fig.~\ref{bievents}, the off-diagonal NSI
	benchmarks shift the predicted event rates toward the observed values,
	allowing both data sets to be simultaneously accommodated.
	
	Throughout this work we adopt the benchmark solutions obtained from
	the combined NO$\nu$A+T2K analysis of
	Ref.~\cite{Chatterjee:2024kbn, Chatterjee:2020kkm}, summarized in
	Table~\ref{tab:benchmarks}. These values lie within the ranges allowed
	by global oscillation analyses including complex NSI couplings
	\cite{Esteban:2018ppq,Esteban:2019lfo}.
	
	Importantly, the same NSI that reconcile the present data modify the
	appearance probability at future experiments. At the longer-baseline
	DUNE experiment the $P_2$ contribution is further enhanced, altering
	both the appearance rate and the CP asymmetry. Consequently, a
	standard three-flavor analysis of DUNE data would misidentify both the
	octant of $\theta_{23}$ and the value of $\dcp$. We now study this
	effect quantitatively.
	
	\begin{table}[t]
		\centering
		\renewcommand{\arraystretch}{1.3}
		\begin{tabular}{lcc}
			\hline
			& Benchmark~1 & Benchmark~2 \\
			\hline
			NSI coupling & $\varepsilon_{e\mu}$ & $\varepsilon_{e\tau}$ \\
			$|\varepsilon|$ & $0.13$ & $0.22$ \\
			$\phi\;[\pi]$ & $1.35$ & $1.70$ \\
			$\sin^2\theta_{23}^{\rm true}$ & $0.56$ & $0.56$ \\
			$\dcp^{\rm true}\;[\pi]$ & $-0.56$ & $-0.58$ \\
			\hline
		\end{tabular}
		\caption{NSI benchmark points adopted in this work, corresponding to
			the best-fit solutions found in a combined analysis of NOvA and T2K
			data under the NO
			hypothesis~\cite{Chatterjee:2024kbn,Chatterjee:2020kkm}.  Both
			points reconcile the NO$\nu$A--T2K tension via off-diagonal
			propagation NSI.}
		\label{tab:benchmarks}
	\end{table}

	\textbf{Implications for DUNE:}
	The Deep Underground Neutrino Experiment (DUNE)~\cite{DUNE:2018tke},
	with a baseline of 1300 km, is designed to disentangle CP violation
	from matter effects and determine the remaining oscillation
	parameters with high precision. Owing to its longer baseline and
	higher neutrino energies compared to T2K and NO$\nu$A, matter
	effects play a significantly larger role.

	Acting as a concrete proxy for generic propagation new 
	physics, we study how the NSI solutions that reconcile the 
	NO$\nu$A–T2K tension would affect the interpretation of 
	DUNE data if analyzed strictly within the standard three-
	flavor framework.
	True spectra are generated with NSI using the benchmark parameters of
	Table~\ref{tab:benchmarks}, while the test spectra assume standard
	oscillations. Simulations are performed with the GLoBES
	framework~\cite{Huber:2004ka,Huber:2007ji} using the official DUNE
	configuration: a 40 kt liquid-argon detector, a 1.2 MW beam at
	120 GeV, and $3.5+3.5$ years of neutrino and antineutrino running,
	corresponding to $\sim300\,\mathrm{kt\cdot MW\cdot yr}$. Systematic
	uncertainties are included through Gaussian pulls following the DUNE
	TDR prescription~\cite{DUNE:2021cuw}. The true oscillation parameters
	are $\sin^2\theta_{23}=0.56$ and $\dcp\simeq-\pi/2$ under NO.
	
	The analysis simultaneously includes $\nue$ ($\anue$) appearance and
	$\numu$ ($\anumu$) disappearance channels. The disappearance channel
	mainly constrains $\sin^22\theta_{23}$ and $|\Delta m_{31}^2|$
	through the survival probability
	$\pmumu\simeq1-\sin^22\theta_{23}\sin^2\dl$, which is symmetric under
	$\theta_{23}\to\pi/2-\theta_{23}$ and therefore insensitive to the
	octant. Since the off-diagonal NSI parameters
	$\varepsilon_{e\mu}$ and $\varepsilon_{e\tau}$ enter the Hamiltonian
	only in the $e\!-\!\mu$ and $e\!-\!\tau$ sectors
	(\cf\ Eq.~\ref{eq:Ham}), their effect on disappearance probabilities
	is small. Consequently the dominant NSI impact arises from the
	appearance channel, leaving the determination of
	$|\Delta m_{31}^2|$ and $\sin^22\theta_{23}$ largely unaffected while
	preserving the octant–CP degeneracy generated by $P_2$.
	
\begin{figure}[t]
		\centering
		\includegraphics[width=\columnwidth]{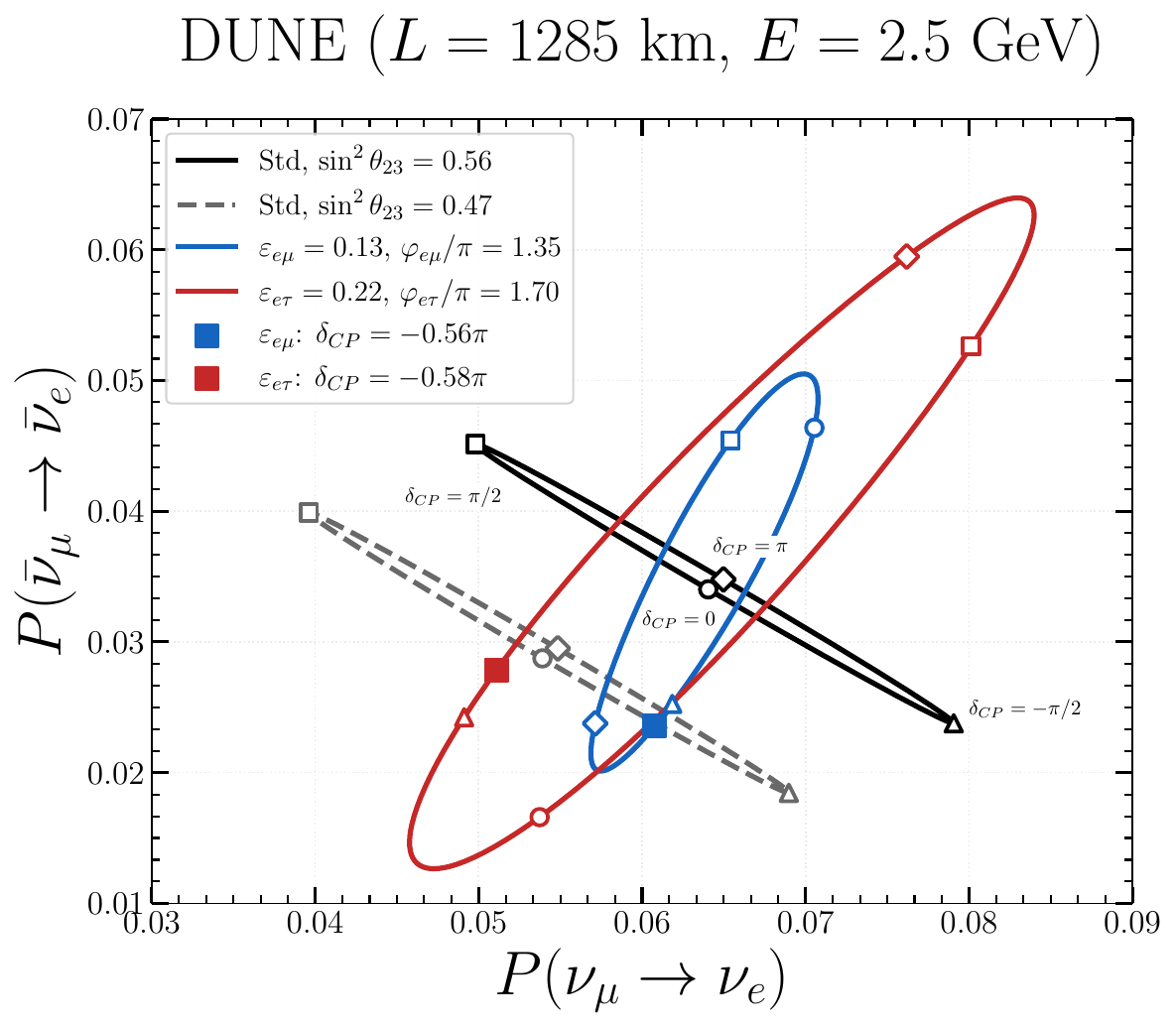}\\[4pt]
		\includegraphics[width=\columnwidth]{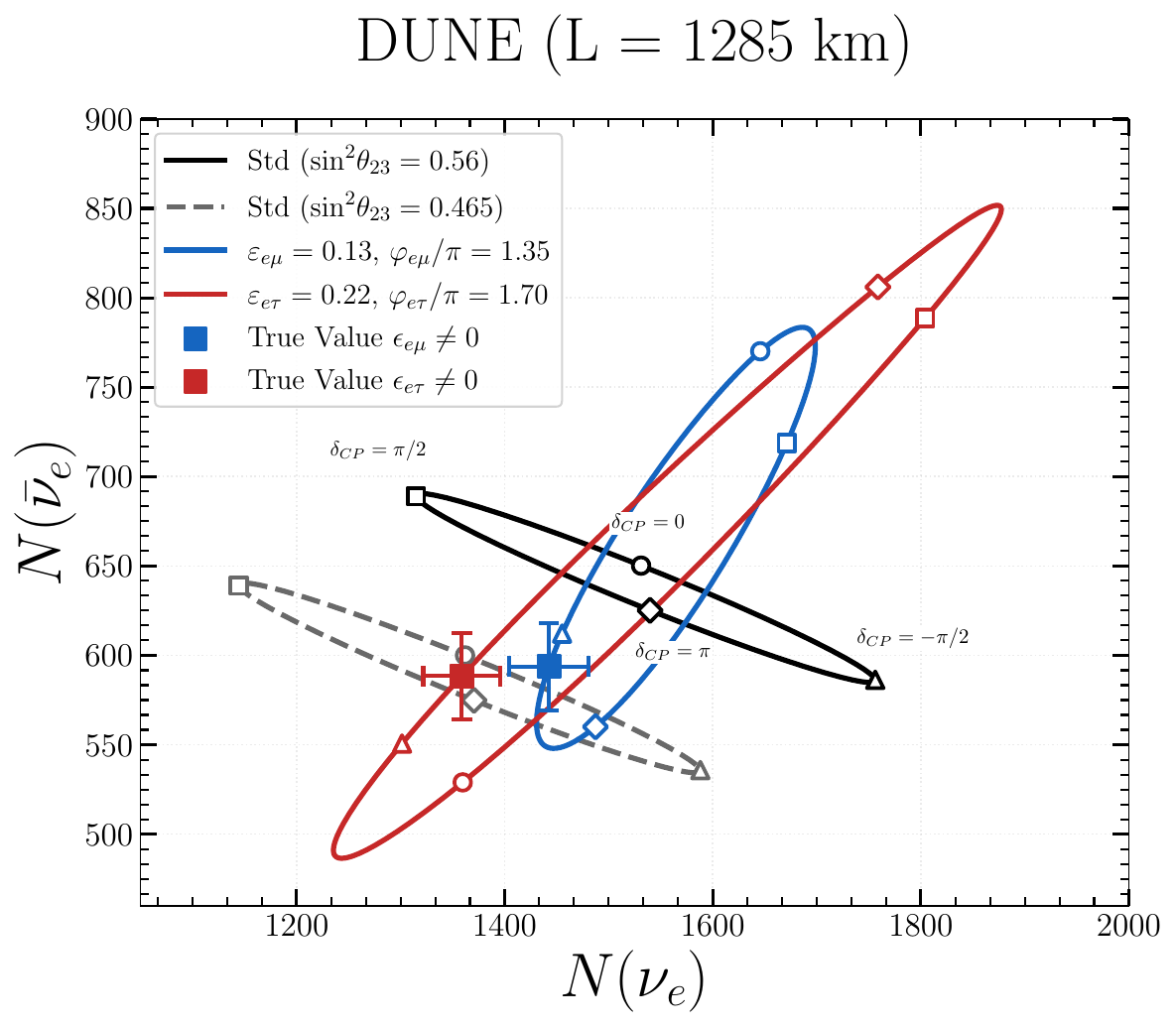}
		\caption{\footnotesize  Bi-probability plot (upper) and  
			bi-event (lower) plot for DUNE. For both cases, the ellipses 
			are obtained by varying $\dcp$ over
			$[-\pi,\pi]$ with all other oscillation parameters fixed at the
			combined best-fit\cite{Esteban:2024eli}.  The predicted probability
			and simulated data points use Benchmarks 1 and 2 
			from table \ref{tab:benchmarks} as references and are represented by 
			colored squares. 
			The black ellipse represents the standard prediction  for the higher 
			octant (HO), while the gray dashed ellipse represents the standard prediction
			for the lower octant (LO). The blue and red ellipses represents the 
			prediction for non-zero NSIs suggested by benchmarks 1 and 2, respectively.
			The circle, square, diamond and triangle shapes represents
			$\dcp=0,\pi/2,\pi(-\pi)$ and $-\pi/2$ respectively.}
		\label{fig:dune-combined}
	\end{figure}

	The upper panel of Fig.~\ref{fig:dune-combined} shows the
	bi-probability plot for DUNE with $\pme$ ($\pmebar$) on the $x$ ($y$)
	axis. For both NSI benchmarks, the NSI ellipse corresponding to
	$\theta_{23}$ in the higher octant overlaps substantially with the
	standard three-flavor ellipse for the lower octant. In
	benchmark~1 the true point lies directly on the LO standard ellipse,
	while in benchmark~2 it lies very close to it. Consequently, event
	rates generated with NSI and $\theta_{23}$ in HO can be reproduced by
	standard oscillations with $\theta_{23}$ in LO and a shifted
	$\dcp$. The lower panel of Fig.~\ref{fig:dune-combined} shows that
	the same degeneracy persists at the level of integrated event
	counts.
	
	Although the analytical discussion below focuses on the oscillation
	maximum, the numerical analysis uses the full DUNE energy spectrum
	($\sim0.5$–$8$ GeV) with the binning specified in the
	TDR~\cite{DUNE:2021cuw}. The bi-event plots incorporate the flux,
	cross sections, and detector response. The close correspondence
	between the bi-probability and bi-event representations indicates
	that the degeneracy persists after spectral integration, since the
	$P_2$ contribution maintains the same sign and magnitude over the
	energy region dominating the event sample.
	
	\afterpage{%
		\begin{figure*}[!t]
			\centering
			\includegraphics[width=\linewidth]{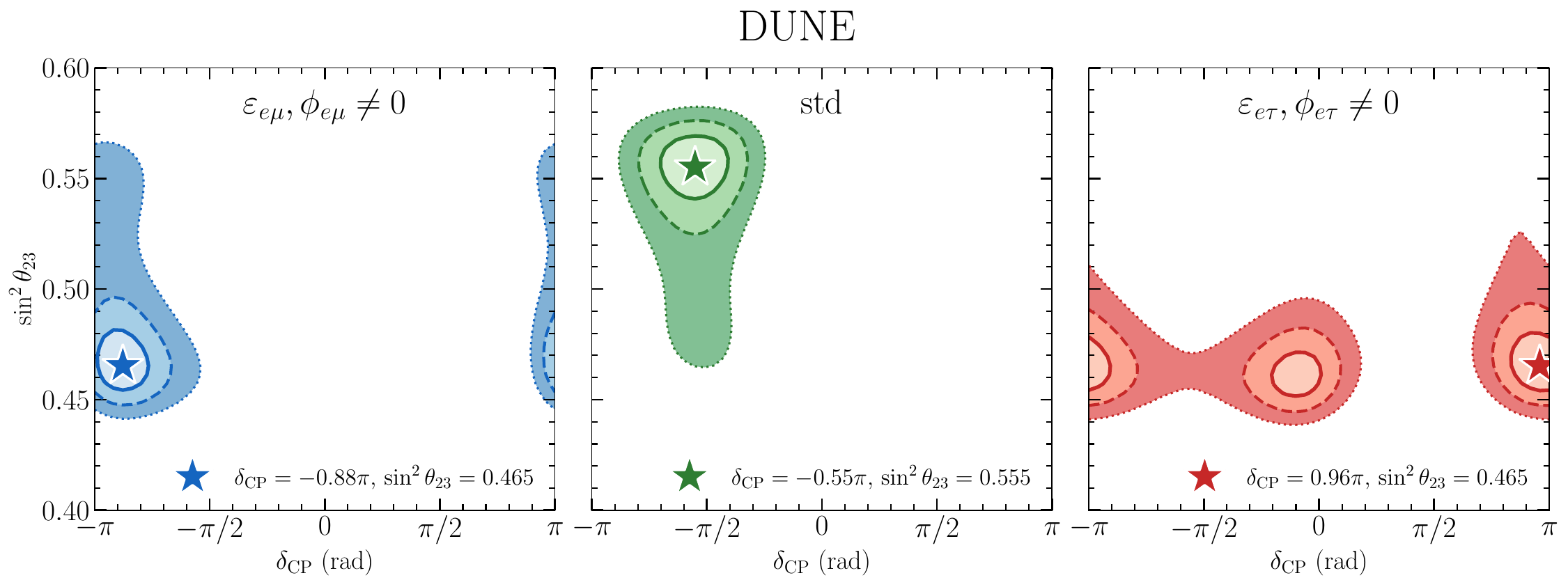}
			\caption{Allowed regions in the $\sin^2\tz$--$\dcp$ plane from a standard
				three-flavor analysis of DUNE data. \emph{Left}: simulated data driven by
				$\varepsilon_{e\mu}$ (benchmark~1); \emph{center}: simulated data driven by
				standard oscillations; \emph{right}: simulated data driven by
				$\varepsilon_{e\tau}$ (benchmark~2). Contours at $1\sigma$, $2\sigma$ and
				$3\sigma$ are shown. The star marks the best-fit point.
			}
			\label{fig:case1_contours}
		\end{figure*}
	}
	Analytically, for DUNE one obtains approximately, for eq.~\ref{P2},
	$\ahat\simeq0.22\,[E/(2.5\,{\rm GeV})]$, where $2.5$~GeV
	corresponds to the first oscillation maximum. At the DUNE flux peak ($\dhat\simeq\pi/2$), the dominant
	CP-odd contribution to the asymmetry
	$A_{CP}=\pme-\pmebar$ arises from $P_1$ (proportional to
	$\sin\dcp$) and from $P_2$ (proportional to
	$\sin(\dcp+\phi)$). The combination $(\dcp+\phi)$ therefore
	acts as an effective CP phase,
	$\dcp^{\rm eff}=\dcp+\phi$. For benchmark~1,
	$\phi_{e\mu}=1.35\pi$ and $\dcp\simeq-\pi/2$ give
	$\dcp^{\rm eff}\simeq0.85\pi$, close to the CP-conserving
	value $\pi$. In this case the NSI contribution suppresses
	the large asymmetry expected at maximal CP violation, and
	a standard oscillation fit compensates by shifting the
	inferred $\dcp$ toward CP-conserving values. For
	benchmark~2 ($\phi_{e\tau}=1.70\pi$),
	$\dcp^{\rm eff}\simeq1.20\pi$ lies in the opposite
	half-plane, driving the fitted $\dcp$ across the
	CP-conserving boundary even though the true value is
	$\dcp\simeq-\pi/2$.
	
	The coefficient $a$ in Eq.~\ref{P2} also determines the
	degree of octant misidentification. For $\varepsilon_{e\mu}$,
	$a=\sin^2\theta_{23}$ distinguishes HO from LO, so the
	NSI ellipse in the bi-probability plane only partially
	overlaps with the LO ellipse of the standard scenario,
	producing a weaker octant degeneracy ($\sim2\,\sigma$).
	In contrast, for $\varepsilon_{e\tau}$,
	$a=s_{23}c_{23}=\tfrac12\sin2\theta_{23}$ is symmetric
	under $\theta_{23}\to\pi/2-\theta_{23}$, implying that
	$P_2$ alone cannot distinguish the two octants. This
	symmetry leads to an intrinsic octant degeneracy when
	both true and test event rates include NSI
	\cite{Agarwalla:2016fkh, Liao:2016hsa}.
	
	Among the three terms in Eq.~\ref{pme-nsi}, the first two
	($P_0$ and $P_1$) reproduce the standard oscillation
	probability $\pme^{\rm std}$. From Eqs.~\ref{P0} and
	\ref{P1}, $\pme^{\rm std}$ is maximized for normal
	ordering, $\dcp=-\pi/2$, and $\theta_{23}$ in the higher
	octant. The additional NSI term $P_2$ reduces this
	probability for the parameter region preferred by the
	combined NO$\nu$A–T2K data, which favor
	$\pi<\phi<2\pi$ at $1\,\sigma$ C.L. In this region
	$\cos(\dcp+\phi)<1$ for $\dcp=-90^\circ$, leading to
	a suppression of the $\nu_e$ (and also $\bar{\nu}_e$) appearance probability. Within the
	standard three-flavor framework this suppression can be
	accommodated by shifting the fitted value of $\theta_{23}$
	from HO to LO and a shift in the fitted value of $\dcp$. Since, the suppression in oscillation probability due to NSI is larger in case of $\varepsilon_{e\tau}$, to accommodate this shift in the fitting with standard oscillation scheme, the fitted $\dcp$ value needs to be shifted to the opposite half plane along with shifting the fitted octant of $\tz$ from HO to LO. Consequently, the wrong-octant solution at
	DUNE arises only for the specific combination of
	$\dcp$ and $\phi$ favored by the NO$\nu$A–T2K data. For any other combination of $\phi$ and $\dcp$, it was possible to remove the wrong octant solution in case of test event generated with standard oscillation scheme.
	
	If a degeneracy occurs between $\pme^{\rm std+NSI}$ in HO
	and $\pme^{\rm std}$ in LO with a shifted CP phase
	$\dcp^\prime$, the condition reads
	\begin{equation}
		(P_0+P_1+P_2)^{\rm std+NSI}_{\rm HO,\phi,\dcp}
		=
		(P_0+P_1)^{\rm std}_{\rm LO,\dcp^\prime}.
		\label{condition}
	\end{equation}
	
	At the DUNE flux peak ($\dhat\simeq\pi/2$), and for the
	special case $\dcp=\dcp^\prime=0$, the $P_1$ term becomes
	negligible on both sides. Taking
	$|\varepsilon_{e\mu}|=0.13$, Eq.~\ref{condition}
	yields $\phi_{e\mu}\simeq0.08\pi$ and $1.03\pi$.
	Similar solutions exist for $\dcp\neq\dcp^\prime$
	and for NSI driven by $\varepsilon_{e\tau}$,
	confirming the generality of the degeneracy.
	
	For antineutrinos the corresponding condition is
	\begin{equation}
		(\bar P_0+\bar P_1+\bar P_2)^{\rm std+NSI}_{\rm HO,\phi,\dcp}
		=
		(\bar P_0+\bar P_1)^{\rm std}_{\rm LO,\dcp^\prime}.
		\label{condition-anu}
	\end{equation}
	Solving Eqs.~\ref{condition} and \ref{condition-anu}
	determines the allowed values of $\phi$ and $\dcp^\prime$, where the $\nu_e$ and $\bar{\nu}_e$ appearance probabilities for NSI due to $\varepsilon_{e\mu}$ ($\varepsilon_{e\tau}$) with a fixed $\phi_{e\mu}$ ($\phi_{e\tau}$) and $\dcp$, with $\tz$ in HO could be mimicked by standard oscillation probability with a $\dcp^\prime$ and $\tz$ in LO.

	The resulting allowed regions in the $\sin^2\theta_{23}$–$\dcp$ plane
	are shown in Fig.~\ref{fig:case1_contours}. In the standard case
	(center panel), DUNE correctly identifies the higher octant at
	$>2\sigma$ and establishes near-maximal CP violation at $>3\sigma$.
	In contrast, when the true spectra contain NSI, the interpretation
	changes qualitatively.
	
	\afterpage{%
		\begin{figure*}[!t]
			\centering
			\includegraphics[width=\linewidth]{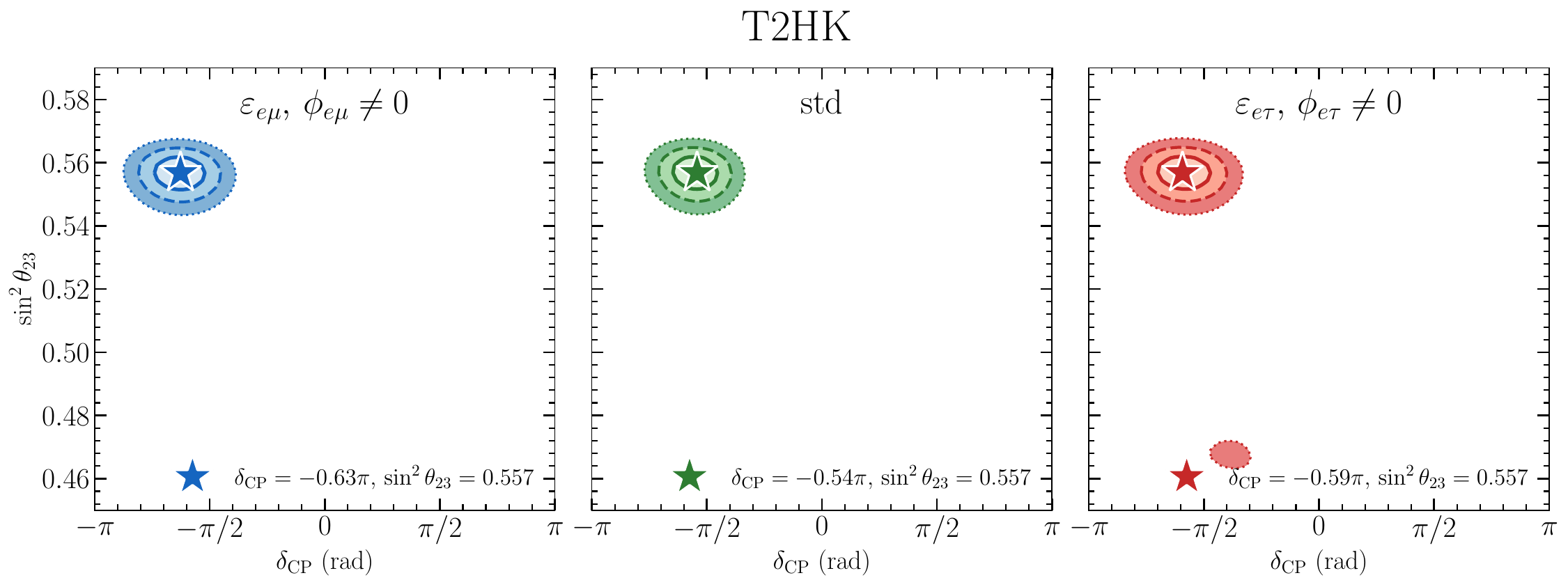}
			\caption{
				Same as Fig.~\ref{fig:case1_contours}, but for a standard
				three-flavor analysis of T2HK data.
			}
			\label{fig:case2_contours}
		\end{figure*}
	}

	For benchmark~1 ($\varepsilon_{e\mu}$), the best-fit value shifts to
	$\dcp\simeq-0.88\pi$, close to CP conservation, while the higher
	octant becomes disfavored at $>2\sigma$. For benchmark~2
	($\varepsilon_{e\tau}$), the distortion is stronger: the lower octant
	is preferred at $>3\sigma$, and the fitted $\dcp$ becomes degenerate
	between $0$ and $\pi$, recovering the true maximal phase only at
	$3\sigma$.
	
	In both cases the biases in $\theta_{23}$ and $\dcp$ arise from the
	same NSI contribution $P_2$, which simultaneously modifies the total
	appearance rate and the CP asymmetry through the effective phase
	$(\dcp+\phi)$. The distinct patterns visible in
	Fig.~\ref{fig:case1_contours} could therefore help identify the
	underlying NSI scenario.
	
	Previous work~\cite{Agarwalla:2016fkh, Blennow:2016etl} showed that NSI can reduce
	DUNE's octant sensitivity when NSI effects are included in both true
	and test spectra. Here we consider the complementary situation in
	which NSI affect the true data but the analysis assumes the standard
	framework. In this case DUNE can incorrectly prefer the \emph{wrong}
	octant. While Ref.~\cite{Liao:2016hsa} predicted such a possibility
	for $\varepsilon_{e\tau}$ with $\dcp=0$, our analysis shows that both
	$\varepsilon_{e\mu}$ and $\varepsilon_{e\tau}$ can induce correlated
	octant and CP-phase misidentification when the true phase is near
	maximal.

	It is worth emphasizing the extent to which this effect is generic.
	The differential impact on DUNE versus T2HK is controlled by the ratio
	of their matter potentials and is therefore present for any non-vanishing
	$|\varepsilon|$: a tension of some magnitude is expected across the
	parameter space, growing with $|\varepsilon|$ and with
	$|\sin(\dcp+\phi)|$. The specific signature studied here---a correlated
	wrong-octant and CP-phase misidentification---is, however, tied to the
	$(|\varepsilon|,\phi)$ region favored by the current NO$\nu$A--T2K fit,
	where the rate suppression $\cos(\dcp+\phi)<1$ is accommodated by an
	HO$\to$LO shift (\cf\ the discussion below Eq.~\ref{condition}). For
	other phase combinations the bias can appear purely as a CP-phase shift
	without octant confusion, or be reabsorbed entirely. A detectable
	DUNE--T2HK tension is thus a broad feature of complex propagation NSI,
	while its precise form is diagnostic of the underlying phase.

	\textbf{T2HK and DUNE–T2HK tension:}
	We repeat the analysis for the Tokai-to-Hyper-Kamiokande (T2HK)
	experiment~\cite{Hyper-KamiokandeProto-:2015xww,Abe:2018uyc,
		Hyper-Kamiokande:2018ofw}.  T2HK will send the upgraded J-PARC beam
	($4$~MW) over a baseline of $L=295$~km to a
	$500$~kt fiducial Water~\v{C}erenkov detector.  We assume an exposure
	of $3\,{\rm yr}\,\nu+3\,{\rm yr}\,\bar\nu$ and simulate the experiment
	using \textsc{GLoBES}~\cite{Huber:2004ka,Huber:2007ji} with the
	official configuration files.  As in the DUNE analysis, the true
	spectra are generated with NSI at the benchmark values of
	Table~\ref{tab:benchmarks}, while the fit assumes standard oscillations.
	The free parameters in the $\chi^2$ minimization are
	$|\datm|$, $\sin^2\tz$, and $\dcp$, with the solar parameters fixed.

	\begin{figure}[!h]
		\centering
		\includegraphics[width=\columnwidth]{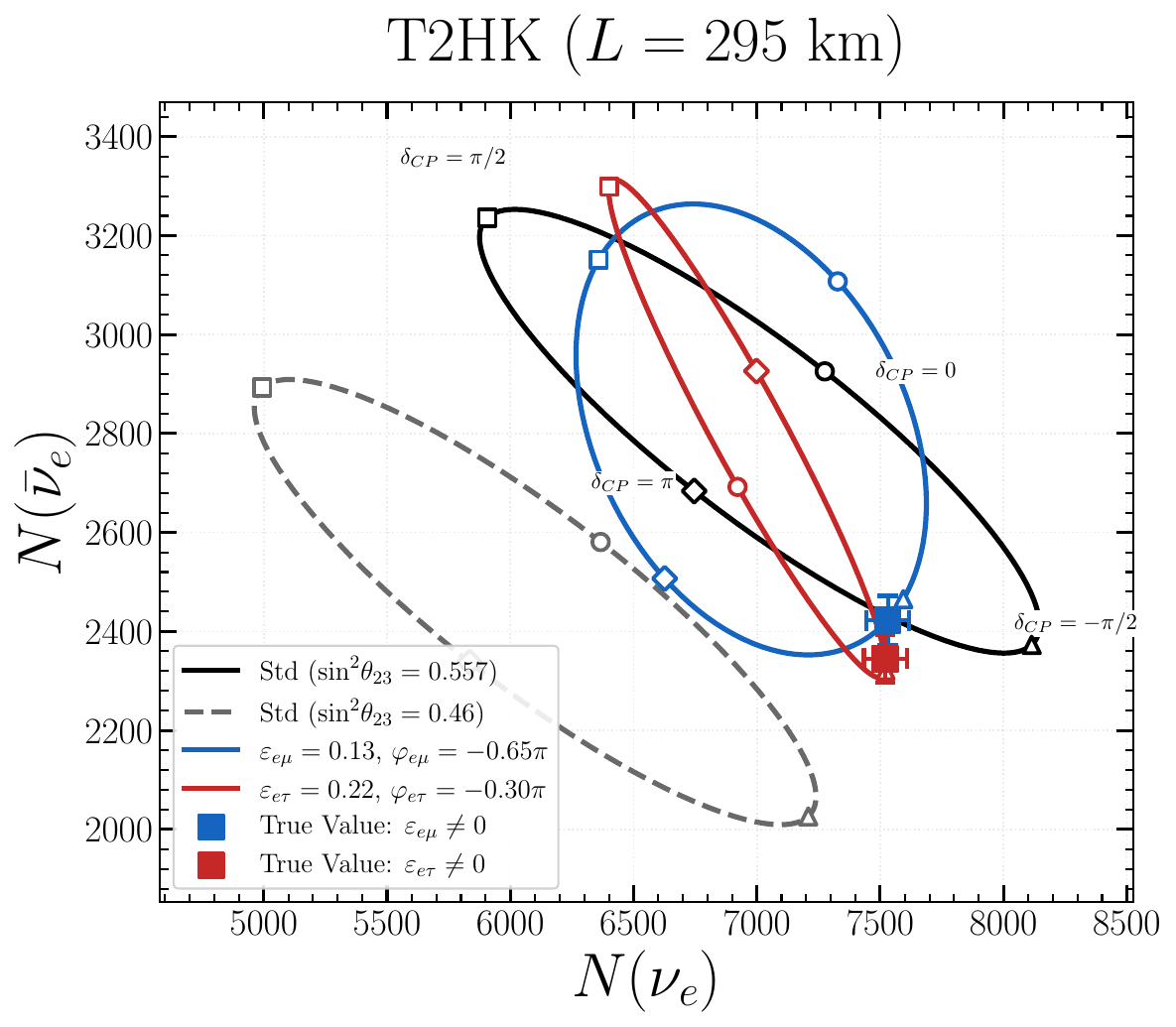}
		\caption{\footnotesize Allowed regions in the $\sin^2\tz$--$\dcp$ plane from a
			standard three-flavor analysis of T2HK data. \emph{Left}: 
			The simulated data with true value driven
			by $\varepsilon_{e\mu}$ (benchmark~1); \emph{center}: the simulated data with
			true value driven standard oscillations; \emph{right}: the simulated data with
			true value driven  by $\varepsilon_{e\tau}$ (benchmark~2).  Contours at
			$1\sigma$, $2\sigma$ and $3\sigma$ are shown.  The star marks the best-fit
			point after analyzing the simulated data using standard oscillation. 
		}
		\label{fig:t2hk-combined}
	\end{figure}
	
	\afterpage{%
		\begin{figure*}[!t]
			\centering
			\includegraphics[width=\linewidth]{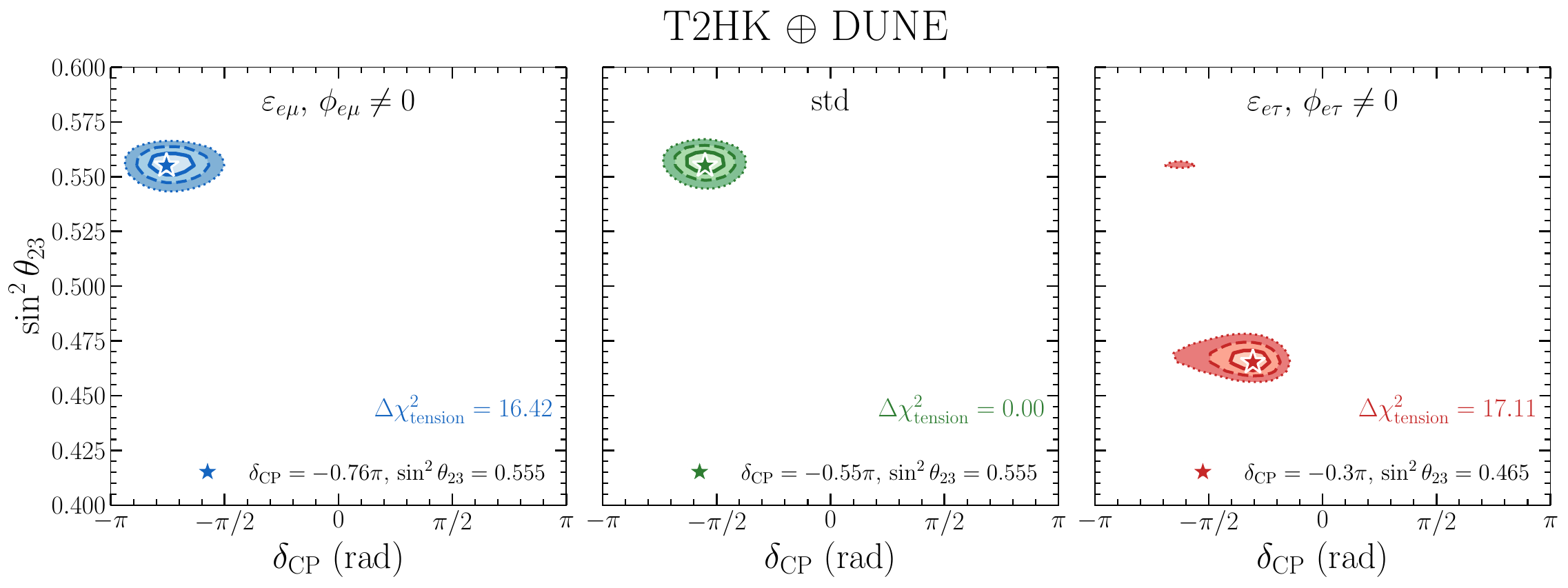}
			\caption{
				Same as Fig.~\ref{fig:case1_contours}, but for a standard
				three-flavor analysis of the combined T2HK\,$\oplus$\,DUNE data. The
				$\Delta\chi^2$ quantifying the tension between the two experiments is
				indicated in each panel.
			}
			\label{fig:case3_contours}
		\end{figure*}
	}
	The shorter baseline of T2HK leads to a much smaller matter potential.
	At the peak energy $E\simeq0.6$~GeV,
	$\ahat\simeq0.05$, about four times smaller than at DUNE
	($\ahat\simeq0.22$ at $E\simeq2.5$~GeV).  Since the NSI term scales
	as $P_2\propto\ahat|\varepsilon|$ [Eq.~(\ref{P2})], the NSI-induced
	shift of the bi-probability ellipse is correspondingly suppressed.
	As illustrated in Fig.~\ref{fig:t2hk-combined}, the NSI and standard
	ellipses lie close to each other, leaving little room for the
	octant–CP degeneracy observed at DUNE.

	The resulting constraints in the $\sin^2\tz$--$\dcp$ plane are shown in
	Fig.~\ref{fig:case2_contours}.  In the standard case (central panel),
	T2HK recovers the higher octant and near-maximal CP violation at high
	significance.  For both NSI benchmarks the fit remains close to the
	true point: the higher octant is preferred at $>2\sigma$ and the true
	phase $\dcp\simeq-0.55\pi$ is recovered within $1$--$2\sigma$.
	The best-fit $\dcp$ shifts only mildly
	($\Delta\dcp\lesssim0.1\pi$), and no degenerate lower-octant solution
	appears at $2\sigma$.
	
	This robustness arises because the smaller matter effect suppresses
	the $P_2$ correction and the large T2HK statistics constrain the fit
	near the true parameters.  The NSI benchmarks therefore do not
	significantly distort the T2HK inference, in sharp contrast to DUNE
	where the larger $P_2$ term drives the fit toward the wrong octant
	and a shifted CP phase.
	
	\textbf{Combined analysis:}
	The contrasting responses of DUNE and T2HK 
	\cite{AbdelKhaleq:2024hir} lead to tension when the
	two data sets are analyzed separately under the standard three-flavor
	hypothesis.  The combined allowed regions are shown in
	Fig.~\ref{fig:case3_contours}.  When the true spectra include NSI,
	DUNE favors a wrong-octant solution with shifted $\dcp$, while T2HK
	recovers the true parameters.  For the combined fit, in case of $\varepsilon_{e\mu}$ ($\varepsilon_{e\tau}$), T2HK\,$\oplus$\,DUNE prefers HO (LO) for $\tz$ with $\dcp$ shifted slightly from the true value.

	To quantify the inconsistency we apply the parameter
	goodness-of-fit (PG) test~\cite{Maltoni:2003cu},
	\begin{equation}
		\chi^2_{\rm PG} =
		\min\!\bigg[\sum_{i}^{N}\chi^2_i\bigg]
		- \sum_{i}^{N}\chi^2_{{\rm min},\,i},
		\label{eq:PG}
	\end{equation}
	which follows a $\chi^2$ distribution with
	$n=\sum_i n_i-n_{\rm glob}$ degrees of freedom, 
	where $n_{\rm glob}$ is the number of parameters varied in common in the
	combined fit.
	We apply this test to the projected
	T2HK and DUNE data sets (appearance and disappearance channels in both
	neutrino and antineutrino modes) under NO, for case 1 (C1): fixing $\dsun$, $\tx$, and
	$\ty$ to their best-fit values, so that each experiment depends on
	three parameters ($|\datm|,\;\sin^2\tz,\;\dcp$), giving $n=3$ degrees
	of freedom, while for case 2 (C2):  fixing $\dsun$ and $\tx$ to their best-fit values, so that each experiment depends on
	three parameters ($|\datm|,\;\sin^2\tz,\;\dcp, \ty$), giving $n=4$ degrees
	of freedom.  The true event spectra are generated with NSI at the
	benchmark values of Table~\ref{tab:benchmarks}, while the fit is
	performed under the standard three-flavor hypothesis.  The results are
	summarized in Table~\ref{tab:PG}.

	\begin{table}[!h]
		\centering
		\renewcommand{\arraystretch}{1.3}
		\begin{tabular}{lccc}
			\hline
			Cases & $\chi^2_{\rm PG}/n$ & $p$-value & $\#\sigma$ \\
			\hline
			C1: Benchmark~1 ($\varepsilon_{e\mu}$) & $16.42/(n=3)$ & $9.3\times10^{-4}$ & $3.3$ \\
			C1: Benchmark~2 ($\varepsilon_{e\tau}$) & $17.11/(n=3)$ & $6.7\times10^{-4}$ & $3.4$ \\
			\hline
			C2: Benchmark~1 ($\varepsilon_{e\mu}$) & $16.42/(n=4)$ & $2.5\times10^{-3}$ & $3.0$ \\
			C2: Benchmark~2 ($\varepsilon_{e\tau}$) & $17.11/(n=4)$ & $1.8\times10^{-3}$ & $3.1$ \\
			\hline
		\end{tabular}
		\caption{Parameter goodness-of-fit test~\cite{Maltoni:2003cu}
			for the consistency of T2HK and DUNE projected data under NO,
			assuming the standard oscillation framework.
			We consider two cases to quantify the tension, in case 1 (C1)
			the $\chi^2_{\rm PG}$ depends on three oscillation parameters.
			In case 2 (C2) the $\chi^2_{\rm PG}$ depends on four oscillation parameters.
			For more details, see the text.}
		\label{tab:PG}
	\end{table}
	
	Both benchmarks yield a T2HK\,$\oplus$\,DUNE incompatibility at the
	$\sim3\sigma$ level when interpreted within the standard
	three-flavor framework after 6 years of data taking,
	increasing the tension with more statistics.
	This arises from the baseline dependence of
	$P_2$: T2HK, largely insensitive to propagation NSI, recovers the
	true oscillation parameters, whereas DUNE infers a shifted $\dcp$
	and the wrong octant.  The resulting discordance therefore provides
	a robust experimental diagnostic of new physics in the neutrino
	propagation sector.
	\\
	
	\textbf{Conclusion:}
	In this letter, we have demonstrated that the baseline complementarity 
	between future long-baseline experiments provides a powerful, 
	generalized diagnostic tool for unmodeled new physics in neutrino propagation. 
	Using the complex off-diagonal NSI---currently favored to resolve the 
	$\sim2\sigma$ NO$\nu$A–T2K tension---as a highly predictive test case, 
	we showed how such beyond Standard Model effects can masquerade as
	standard parameter degeneracies.
	
	Because matter effects are significantly larger at DUNE, a standard 
	three-flavor analysis of its data can simultaneously misidentify 
	the octant of $\theta_{23}$ and the leptonic CP phase $\dcp$. 
	For $\varepsilon_{e\mu}$ NSI, the inferred $\dcp$ shifts toward 
	CP-conserving values while the higher octant becomes disfavored at 
	$2\,\sigma$, whereas for $\varepsilon_{e\tau}$ NSI, the fitted 
	$\dcp$ can flip to the opposite half-plane together with a 
	strong preference for the lower octant higher than $3\,\sigma$. 
	Both effects arise because the complex NSI phase $\phi$ modifies 
	the CP asymmetry through the effective combination $\dcp+\phi$, 
	leading to a large suppression of $\nu_e$ and $\bar{\nu}_e$ 
	appearance probabilities that the standard fit erroneously 
	accommodates by shifting to the lower octant. In contrast, 
	T2HK is largely insensitive to these propagation effects owing 
	to its shorter baseline, correctly recovering the true oscillation parameters.
	
	The resulting mismatch between the DUNE and T2HK standard-fit 
	results produces a $3.3$–$3.4\sigma$ inconsistency in a parameter 
	goodness-of-fit test after 6 years of data collecting, 
	getting increased with more statistics. 
	This predicted tension between DUNE and T2HK 
	is directly analogous to the current NO$\nu$A–T2K discrepancy: 
	in both cases, the tension arises because unmodeled propagation 
	physics has a substantially stronger impact on the longer-baseline experiment. 
	
	While we have utilized NSI to quantitatively illustrate this effect, 
	our findings establish a broader principle. If nature contains new physics 
	in the neutrino propagation sector, assuming the standard hypothesis 
	can force experiments with distinct matter effects to infer divergent 
	parameters. Baseline complementarity between DUNE and T2HK therefore can provides 
	a robust and generic diagnostic: a future tension between their measurements of 
	the octant or CP phase would be a compelling indication of BSM physics, 
	with the current NO$\nu$A–T2K discrepancy serving as a clear prelude 
	to the degeneracies we must be prepared to disentangle.

	\textbf{Acknowledgment:}
	We thank Pedro Machado for useful insights.
	JPP is supported by the National Natural Science Foundation of China (12425506 and 12375101). This work is also supported by State Key Laboratory of Dark Matter Physics.  U.R.\ would like to
	acknowledge support from the Department of Atomic Energy, Government
	of India (Project Identification Number RTI 4002) and from the J.~C.\
	Bose Grant of the Anusandhan National Research Foundation (ANRF),
	Government of India (ANRF/JBG/2025/000265/PS).
	UR dedicates this letter to Dr.\ Umar Khalid and Sharjeel Imam, who
	have been imprisoned in India without trial for more than five years.
	Their courage in standing by their convictions, even in the face of
	severe personal consequences, serves as a reminder of the importance
	of speaking truth to power.  UR emphasizes that this dedication
	reflects solely his personal views.  Neither the co-author of this
	paper nor the Tata Institute of Fundamental Research bears any
	responsibility for this statement.

	\begin{widetext}
		\bibliographystyle{apsrev}
		\bibliography{references}
	\end{widetext}

\end{document}